\documentclass[11pt,twoside]{article}
\usepackage[pdftex]{graphicx}
\usepackage{amsmath}
\usepackage{amssymb}
\usepackage{cite}

\begin{document}
\newcommand{\pst}{\hspace*{1.5em}}

\newcommand{\rigmark}{\em Journal of Russian Laser Research}
\newcommand{\lemark}{\em Volume 30, Number 5, 2009}

\newcommand{\be}{\begin{equation}}
\newcommand{\ee}{\end{equation}}
\newcommand{\bm}{\boldmath}
\newcommand{\ds}{\displaystyle}
\newcommand{\bea}{\begin{eqnarray}}
\newcommand{\eea}{\end{eqnarray}}
\newcommand{\ba}{\begin{array}}
\newcommand{\ea}{\end{array}}
\newcommand{\arcsinh}{\mathop{\rm arcsinh}\nolimits}
\newcommand{\arctanh}{\mathop{\rm arctanh}\nolimits}
\newcommand{\bc}{\begin{center}}
\newcommand{\ec}{\end{center}}

\thispagestyle{plain}

\label{sh}


\begin{center} {\Large \bf
\begin{tabular}{c}
THE MAPS OF MATRICES
\\[-1mm]
AND PORTRAIT MAPS OF DENSITY OPERATORS\\[-1mm]
OF COMPOSITE AND NONCOMPOSITE SYSTEMS
\end{tabular}
 } \end{center}

\bigskip

\bigskip

\begin{center} {\bf
Margarita A. Man'ko$^{1*}$ and Vladimir I. Man'ko$^{1,\,2}$
}\end{center}

\medskip

\begin{center}
{\it $^1$P.~N.~Lebedev Physical Institute, Russian Academy of Sciences\\
Leninskii Prospect 53, Moscow 119991, Russia

\smallskip

$^2$Moscow Institute of Physics and Technology (State University)\\
Institutski\'{\i} per. 9, Dolgoprudny\'{\i}, Moscow Region 141700,
Russia}

\smallskip

$^*$Corresponding author e-mail:~~~mmanko~@~sci.lebedev.ru\\
\end{center}

\begin{abstract}\noindent
We obtain a new inequality for arbitrary Hermitian matrices. We
describe particular linear maps called the matrix portrait of
arbitrary $N$$\times$$N$ matrices. The maps are obtained as analogs
of partial tracing of density matrices of multipartite qudit
systems. The structure of the maps is inspired by ``portrait'' map
of the probability vectors corresponding to the action on the
vectors by stochastic matrices containing either unity or zero
matrix elements. We obtain new entropic inequalities for arbitrary
qudit states including a single qudit and an discuss entangled
single qudit state. We consider in detail the examples of $N=3$ and
$4$. Also we point out a possible use of entangled states of systems
without subsystems (e.g., a single qudit) as a resource for quantum
computations.

\end{abstract}

\medskip

\noindent{\bf Keywords:} quantum channels, positive maps, partial trace,
quantum entanglement, nonseparable states.

\section{Introduction}
\pst
The quantum states of arbitrary systems including the spin (qudit)
systems are identified with the density matrices
$\rho$~\cite{Landau27,vonNeumann,vonNeumann1}, which are Hermitian
nonnegative matrices $\rho=\rho^\dagger$ and $\rho\geq 0$ with
Tr$\,\rho=1$. The influence of different devices and measurements on
the system state is associated with a map of the density matrix
$\rho\to\rho'=\Phi(\rho)$, where the matrix $\rho'$ belongs to the
set of density matrices. If the map is a linear transform of the
density matrix, it is called the positive map. Mathematical and
physical aspects of the positive maps have been discussed
in~\cite{Stinospring,SudMatRaoPhysRev1961}.

Particular positive maps corresponding to the transformation of the
density matrix of a physical system inspired by the interaction of
the system with environment are called the completely positive maps.
A set of completely positive maps is the subset of positive maps. In
quantum information theory, the completely positive maps are called
the quantum channels~\cite{Holevobook}. If the function $\Phi(\rho)$
does not describe the linear transform of the density matrix, the
map is called the nonlinear map or the nonlinear quantum
channel~\cite{Puzko}. In the probability representation of quantum
states~\cite{ManciniPLA96,OlgaJETP,DodPLA,VentriPS,NuovoCim}, the
examples of nonlinear positive maps were given in \cite{MMSV} on the
example of unitary tomograms of qudit states.

In composite quantum systems, the entanglement
phenomenon~\cite{Schroed35} corresponding to strong quantum
correlations between the subsystems takes place. In some cases, the
entanglement can be detected applying the portrait map of the
density matrix of the composite-system
state~\cite{VovaJRLR,LupoJPA}. The portrait of the density matrix is
a very particular example of the positive maps of the density
matrix. The common properties of the probability vectors within the
classical and quantum frameworks were studied in \cite{JRLR2014}.
Recently, it was observed that the properties of quantum
entanglement and other aspects of quantum
correlations~\cite{OlgaVova,OlgaVova1,RitaPS2014,OlgaVovaVI,Puzko,Markovich}
existing in composite quantum systems, e.g., in the form of entropic
inequalities~\cite{Ruakai1,Ruakai2,Mendes,Bregenz,RitaPS2014,Petz},
exist also in a single qudit system.

In fact, the mathematical structure for formulating the quantum
correlation properties of composite systems in the form of
equalities and inequalities for the density matrices of such systems
can be found for the density matrix of the systems without
subsystems as well. This fact provides the possibility to obtain the
entanglement properties and new entropic inequalities for density
matrices of the systems without subsystems. One can also formulate
all Bell-like inequalities~\cite{SHRS,AkopyanJRLR,GisinArXiv} and
study the violation of these inequalities for the systems without
subsystems.

It is assumed that the resource for developing fast quantum
computations is associated with the properties of entangled states
of composite quantum systems~\cite{Nielson}. For example, the system
of $N$ qubits is such a system. From the observation above mentioned
follows that an analogous resource can be associated with entangled
states of a single qudit with large spin $j$, such that $2j+1=2^N$.
So, for $N=2$ this means that the system of two qubits, i.e., two
spins with $j=1/2$, has the same entanglement resource as one qudit
with $j=3/2$.

The aim of our work is to show that any $N$$\times$$N$ density
matrix $\rho$ of a quantum system satisfies the same entropic
inequalities associated with the properties of different matrices
$\Phi_1(\rho),\Phi_2(\rho),\dots,\Phi_n(\rho)$, either being
identified or independently of the identification of the matrices
$\Phi_k(\rho)$ with the density matrices of the subsystem states of
the system under consideration. From mathematical point of view, our
goal is to present an analog of the entropic inequality for an
arbitrary Hermitian matrix that seems to be a new inequality in
matrix theory.

This new inequality, being applied to nonnegative trace-class
Hermitian matrices, provides a new entropic inequality for
quantum-state density matrices. The structure of the inequality
makes clear the possibility to introduce the notion of entanglement
and other quantum correlation aspects as characte\-ristics of both
composite and noncomposite quantum systems. This idea is coherent
with the approach to hidden variables for spin $j=1$
state~\cite{Klyachko}. Since the entanglement phenomenon
corresponding to quantum correlations of subsystems of composite
systems provides a resource for quantum computations, we also
consider the possibility to use the entangled states of noncomposite
systems as such a resource as well.

This paper is organized as follows.

In Sec.~2, we give the explanation of new inequalities valid for
arbitrary Hermitian matrices. In Sec.~3, we consider a few examples
of matrix inequalities for three-dimensional and four-dimensional
matrices. In Sec.~4, we present our conclusions and the
prospectives.

\section{Linear Map of
{$\boldsymbol{N\!\!\times\!\!N}$} Matrices}
\pst
The $N$$\times$$N$ matrix $A$ with matrix elements $A_{jk}$ can be
mapped onto the $N^2$-vector $\vec A$. For example, for $N=2$, $A=
\left(\begin{array}{cc}
A_{11}&A_{12}\\
A_{21}&A_{22}\end{array}\right)$, and we define the vector $\vec A$
as the column vector  $\vec A=
\left(\begin{array}{c}
A_{11}\\A_{12}\\A_{21}\\A_{22}\end{array}\right);$ the map is
invertible.

For the map of indices $11\leftrightarrow 1$, $12\leftrightarrow 2$,
$21\leftrightarrow 3$, and $22\leftrightarrow 4$, the vector $\vec
A$ has four components $\vec A=(A_1,A_2,A_3,A_4)$. We can define an
analogous map for an arbitrary $N$.

The linear map of matrices $A\to A'$ can be considered as a map of
vectors, i.e., $A_{jk}\to A'_{jk}=\sum_{m,n=1}^NB_{jk,\,mn}A_{mn}$
can be considered as the relation $A_\alpha\to
A'_{\alpha}=\sum_{\beta=1}^{N^2}b_{\alpha,\,\beta}A_{\beta}$.

For nonnegative Hermitian matrices $A$, such that Tr$\,A=1$ and the
eigenvalues of the matrix are nonnegative, the linear map $A\to A'$
is called the positive map, if the matrices $A'$ have the same
properties. The structure of matrices $b_{\alpha,\,\beta}$ and
$B_{jk,\,mn}$ has been studied in \cite{SudMatRaoPhysRev1961}.

In this paper, we consider specific linear maps of arbitrary complex
matrices $A$.

Let $N=nm$ with $n$ and $m$, the integers. There exist two matrices
$A_1$ and $A_2$ obtained by applying the linear maps, $A\to A_1$ and
$A\to A_2$. We define the maps as follows.

Let the matrix $A$ be presented in the block form
\begin{equation}\label{LM1}
A=\left(\begin{array}{cccc}
a_{11}&a_{12}&\cdots&a_{1n}\\
a_{21}&a_{22}&\cdots&a_{2n}\\
\cdots &\cdots &\cdots&\cdots\\
a_{n1}&a_{n2}&\cdots&a_{nn}\end{array}\right),
\end{equation}
where the blocks $a_{jk}$ $(j,k=1,2,\ldots,n)$ are the
$m$$\times$$m$ matrices. The maps we defined read
\begin{equation}\label{LM2}
A\to A_1=\left(\begin{array}{cccc}
\mbox{Tr}\,a_{11}&\mbox{Tr}\,a_{12}&\cdots&\mbox{Tr}\,a_{1n}\\
\mbox{Tr}\,a_{21}&\mbox{Tr}\,a_{22}&\cdots&\mbox{Tr}\,a_{2n}\\
\cdots &\cdots &\cdots&\cdots\\
\mbox{Tr}\,a_{n1}&\mbox{Tr}\,a_{n2}&\cdots&\mbox{Tr}\,a_{nn}\end{array}\right),\qquad
A\to A_2=\sum_{k=1}^na_{kk}=\sum_{k,j=1}^na_{kj}\delta_{kj}.
\end{equation}
Thus, we obtained two matrices: the $n$$\times$$n$ matrix $A_1$ and
the $m$$\times$$m$ matrix $A_2$. The constructed map preserves the
trace, i.e., Tr$\,A=\mbox{Tr}\,A_1=\mbox{Tr}\,A_2$, and if
$A^\dagger=A$, then $A_1^\dagger=A_1$ and $A_2^\dagger=A_2$. The map
$A\to A_1\otimes A_2$ is the nonlinear map.

The map constructed has the invariance properties. If we replace the
matrix $A$ by the matrix
$$A_u=(1_n\otimes u_m)A(1_n\otimes u_m^\dagger),$$
where $u_m$ is the unitary $m$$\times$$m$ matrix, and $1_n$ is the
$n$$\times$$n$ identity matrix, the matrix $A_{1u}$ (obtained by the
described procedure from the matrix $A_u$) does not depend on the
unitary matrix $u_n$, i.e., $A_{1u}=A_1$.

Analogously, if the matrix $A$ is replaced by the matrix
$$\widetilde A_u=(u_n\otimes 1_m)A(u_n^\dagger\otimes 1_m),$$
one has the property $A_{2u}=A_2$. We can prove that for an
arbitrary Hermitian $N$$\times$$N$ matrix $A=A^\dagger$, such that
Tr$\,A=1$ and $A\geq 0$, the inequality $~-\mbox{Tr}\,A\ln A\leq
-\mbox{Tr}\,A_1\ln A_1-\mbox{Tr}\,A_2\ln A_2~$ is valid.

If Tr$\,A^2=\mbox{Tr}\,A=1$ (pure state), the set of eigenvalues of
$A_1$ and $A_2$ is the same set. It is true for an arbitrary
possible factorization $N=nm=n'm'$. With the matrix $A$, one can
associate an analog of the mutual information with respect to the
decomposition $N=nm$. We introduce the mutual matrix information
\begin{equation}\label{LM3}
I_{nm}=-\mbox{Tr}\,A_1\ln A_1-\mbox{Tr}\,A_2\ln A_2+\mbox{Tr}\,A\ln
A.
\end{equation}
If we introduce $N$$\times$$N$ matrices $\widetilde A_1$ and
$\widetilde A_2$,
\begin{equation}\label{LM4}
\widetilde A_1=\left(\begin{array}{ccc}
A_1&0\\
0&0\end{array}\right),\qquad
\widetilde A_2=\left(\begin{array}{ccc}
A_2&0\\
0&0\end{array}\right),
\end{equation}
the mutual information is determined as
\begin{equation}\label{LM5}
I_{nm}=\mbox{Tr}\left(A\ln A-\widetilde A_1\ln
\widetilde A_1-\widetilde A_2\ln\widetilde A_2\right)\geq 0.
\end{equation}

Now we are in a position to formulate a new statement on the
properties of matrices.

Given $N$$\times$$N$ matrix $A$ [Eq.~(\ref{LM1})] such that
$A=A^\dagger$, $A\geq 0$, and Tr$\,A=1$, this matrix is presented in
the block form with $n^2$ block $m$$\times$$m$ matrices $a_{jk}$
having matrix elements $(a_{jk})_{\alpha\beta}$
$\alpha,\beta=1,2,\ldots,m$, and $N=nm$, then the inequality holds
\begin{eqnarray}\label{LM6}
&&~~-\mbox{Tr}\left\{\left(\begin{array}{cccc}
a_{11}&a_{12}&\cdots&a_{1n}\\
a_{21}&a_{22}&\cdots&a_{2n}\\
\cdots &\cdots &\cdots&\cdots\\
a_{n1}&a_{n2}&\cdots&a_{nn}\end{array}\right)\ln\left(\begin{array}{cccc}
a_{11}&a_{12}&\cdots&a_{1n}\\
a_{21}&a_{22}&\cdots&a_{2n}\\
\cdots &\cdots &\cdots&\cdots\\
a_{n1}&a_{n2}&\cdots&a_{nn}\end{array}\right)\right\}\nonumber\\
&&\leq -\mbox{Tr}\left\{\left(\begin{array}{cccc}
\mbox{Tr}\,a_{11}&\mbox{Tr}\,a_{12}&\cdots&\mbox{Tr}\,a_{1n}\\
\mbox{Tr}\,a_{21}&\mbox{Tr}\,a_{22}&\cdots&\mbox{Tr}\,a_{2n}\\
\cdots &\cdots &\cdots&\cdots\\
\mbox{Tr}\,a_{n1}&\mbox{Tr}\,a_{n2}&\cdots&\mbox{Tr}\,a_{nn}\end{array}\right)\ln
\left(\begin{array}{cccc}
\mbox{Tr}\,a_{11}&\mbox{Tr}\,a_{12}&\cdots&\mbox{Tr}\,a_{1n}\\
\mbox{Tr}\,a_{21}&\mbox{Tr}\,a_{22}&\cdots&\mbox{Tr}\,a_{2n}\\
\cdots &\cdots &\cdots&\cdots\\
\mbox{Tr}\,a_{n1}&\mbox{Tr}\,a_{n2}&\cdots&\mbox{Tr}\,a_{nn}\end{array}\right)\right\}\nonumber\\
&&~~-\mbox{Tr}\left\{(a_{11}+a_{22}+\cdots
+a_{nn})\ln(a_{11}+a_{22}+\cdots +a_{nn})\right\}.
\end{eqnarray}
At $N\neq nm$, we choose an integer $s$, such that the number
$\widetilde N=N+s=nm$, and construct the $\widetilde
N$$\times$$\widetilde N$ matrix
\begin{equation}\label{LM4a}
\widetilde A=\left(\begin{array}{cc}
A&0\\
0&0\end{array}\right).
\end{equation}
After presenting the matrix $\widetilde A$ in an analogous block
form and taking into account zero matrix elements, we arrive at the
inequality
\begin{equation}\label{LM7}
-\mbox{Tr}\,(\widetilde A\ln \widetilde A)=-\mbox{Tr}\,(A\ln A)\leq
-\mbox{Tr}\,(A_1\ln A_1)-\mbox{Tr}\,(A_2\ln A_2).
\end{equation}
At Tr$\,A=\mu$ and $A\geq 0$, we obtain an analogous inequality,
where the term $\mu\ln\mu$ is taken into account; the inequality
reads
\begin{equation}\label{LM8} -\mbox{Tr}\,(A\ln A)\leq
-\mbox{Tr}\,(A_1\ln A_1)-\mbox{Tr}\,(A_2\ln
A_2)+\mbox{Tr}\,A\ln(\mbox{Tr}\,A).
\end{equation}
If the $N$$\times$$N$ matrix $A$ and the $\widetilde
N$$\times$$\widetilde N$ matrix $\widetilde A$ with
Tr$\,A=\mbox{Tr}\,\widetilde A=\mu$, have the minimum eigenvalue
$A_0$, which can be either positive or negative, then for an
arbitrary positive number $x\geq|A_0|$ the matrix $A'(x)=\widetilde
A+x1_{\widetilde N}\geq 0$, and for this matrix we obtain the
inequality
\begin{equation}\label{LM9}
-\mbox{Tr}\,\big(A'(x)\ln A'(x)\big)\leq -\mbox{Tr}\,\big(A'_1(x)\ln
A'_1(x)\big)-\mbox{Tr}\,\big(A'_2(x)\ln
A'_2(x)\big)+\big(\mbox{Tr}\,A'(x)\big)\big(\ln\mbox{Tr}\,A'(x)\big).
\end{equation}
Here, the $\widetilde N$$\times$$\widetilde N$ matrix $\widetilde A$
is expressed in terms of the $N$$\times$$N$ matrix $A$ by
Eq.~(\ref{LM4a}), the integer $\widetilde N=n(\widetilde N/n)$,
where the integer $\widetilde N/n=m$, and the $\widetilde
N$$\times$$\widetilde N$ matrix $A'(x)=\widetilde A+x1_{\widetilde
N}$, where $1_{\widetilde N}$ is the identity matrix in the
$\widetilde N$-dimensional space.

In an explicit form, inequality~(\ref{LM9}) reads
\begin{eqnarray}\label{LM10}
-\mbox{Tr}\left\{\left(\begin{array}{cccc}
a_{11}+x1_m&a_{12}&\cdots&a_{1n}\\
a_{21}&a_{22}+x1_m&\cdots&a_{2n}\\
\cdots &\cdots &\cdots&\cdots\\
a_{n1}&a_{n2}&\cdots&a_{nn}+x1_m\end{array}\right)\ln
\left(\begin{array}{cccc}
a_{11}+x1_m&a_{12}&\cdots&a_{1n}\\
a_{21}&a_{22}+x1_m&\cdots&a_{2n}\\
\cdots &\cdots &\cdots&\cdots\\
a_{n1}&a_{n2}&\cdots&a_{nn}+x1_m\end{array}\right)\right\}\nonumber\\
-(\widetilde Nx+\mbox{Tr}\,A)\ln(\widetilde
Nx+\mbox{Tr}\,A) 
\leq -\mbox{Tr}\left\{\left(\begin{array}{cccc}
\mbox{Tr}\,(a_{11}+x1_m)&\mbox{Tr}\,a_{12}&\cdots&\mbox{Tr}\,a_{1n}\\
\mbox{Tr}\,a_{21}&\mbox{Tr}\,(a_{22}+x1_m)&\cdots&\mbox{Tr}\,a_{2n}\\
\cdots &\cdots &\cdots&\cdots\\
\mbox{Tr}\,a_{n1}&\mbox{Tr}\,a_{n2}&\cdots&\mbox{Tr}\,(a_{nn}+x1_m)\end{array}\right)\right.\nonumber\\
\left.\times\ln\left(\begin{array}{cccc}
\mbox{Tr}\,(a_{11}+x1_m)&\mbox{Tr}\,a_{12}&\cdots&\mbox{Tr}\,a_{1n}\\
\mbox{Tr}\,a_{21}&\mbox{Tr}\,(a_{22}+x1_m)&\cdots&\mbox{Tr}\,a_{2n}\\
\cdots &\cdots &\cdots&\cdots\\
\mbox{Tr}\,a_{n1}&\mbox{Tr}\,a_{n2}&\cdots&\mbox{Tr}\,(a_{nn}+x1_m)\end{array}\right)\right\}\nonumber\\
~\hspace{-40mm}-\mbox{Tr}\left\{(a_{11}+a_{22}+\cdots+a_{nn}+nx1_m)\ln(a_{11}+a_{22}+\cdots+a_{nn}+nx1_m)
\right\},\nonumber\\
\end{eqnarray}
where $1_m$ is the identity matrix in the $m$-dimensional space.

For $N=\widetilde N$, $\widetilde A=A$, and if $x=0$ and $A\geq 0$,
Eq.~(\ref{LM10}) converts in Eq.~(\ref{LM6}), if Tr$\,A=1$.

The matrix entropic inequality~(\ref{LM10}) is the main new relation
found in our work. It can be applied to the ``separable'' matrix
$A$, which has the form of convex sum $A=\sum_kp_kA_k^{(1)}\otimes
A_k^{(2)}$, or to the entangled matrix $A$. If the matrix $A$ is the
diagonal one, inequality~(\ref{LM10}) is the inequality for real
vectors and, if the components of the vectors are nonnegative, we
have the entropic inequalities for the probability vectors.

\section{Examples of $\boldsymbol{\it N}=\,$3 and 4}
\pst
In this scetion, we present the inequalities for particular values
of $N$. We consider the Hermitian 4$\times$4 matrix $A$ given in the
block form $(4=2\cdot 2)$, i.e., $m=2$ and $n=2$,
$~A=\left(\begin{array}{cc}
a_{11}&a_{12}\\
a_{21}&a_{22}\end{array}\right)$, where
\begin{equation}\label{EX1}
a_{11}=
\left(\begin{array}{cc}
\rho_{11}&\rho_{12}\\
\rho_{21}&\rho_{22}\end{array}\right),\quad
a_{12}= \left(\begin{array}{cc}
\rho_{13}&\rho_{14}\\
\rho_{23}&\rho_{24}\end{array}\right),\quad
a_{21}=
\left(\begin{array}{cc}
\rho_{31}&\rho_{32}\\
\rho_{41}&\rho_{42}\end{array}\right),\quad
a_{22}=
\left(\begin{array}{cc}
\rho_{33}&\rho_{34}\\
\rho_{43}&\rho_{44}\end{array}\right).\end{equation}
We have $~\mbox{Tr}\,a_{11}=\rho_{11}+\rho_{22}$,
$~\mbox{Tr}\,a_{12}=\rho_{13}+\rho_{24}$,
$~\mbox{Tr}\,a_{21}=\rho_{41}+\rho_{42},~$ and
$~\mbox{Tr}\,a_{22}=\rho_{33}+\rho_{44}$.

Let $A=A^\dagger$,
~Tr$\,A=\mu=\rho_{11}+\rho_{22}+\rho_{33}+\rho_{44},~$ and $A\geq
0$. In this case, we have the inequality
\begin{eqnarray}\label{EX2}
&&~~-\mbox{Tr}\left\{
\left(\begin{array}{cccc}
\rho_{11}&\rho_{12}&\rho_{13}&\rho_{14}\\
\rho_{21}&\rho_{22}&\rho_{23}&\rho_{24}\\
\rho_{31}&\rho_{32}&\rho_{33}&\rho_{34}\\
\rho_{41}&\rho_{42}&\rho_{43}&\rho_{44}
\end{array}\right)\ln
\left(\begin{array}{cccc}
\rho_{11}&\rho_{12}&\rho_{13}&\rho_{14}\\
\rho_{21}&\rho_{22}&\rho_{23}&\rho_{24}\\
\rho_{31}&\rho_{32}&\rho_{33}&\rho_{34}\\
\rho_{41}&\rho_{42}&\rho_{43}&\rho_{44}
\end{array}\right)\right\}\nonumber\\
&&~~-(\rho_{11}+\rho_{22}+\rho_{33}+\rho_{44})\ln
(\rho_{11}+\rho_{22}+\rho_{33}+\rho_{44})\nonumber\\
&&\leq-\mbox{Tr}\left\{
\left(\begin{array}{cc}
\rho_{11}+\rho_{22}&\rho_{13}+\rho_{24}\\
\rho_{31}+\rho_{42}&\rho_{33}+\rho_{44}
\end{array}\right)\ln
\left(\begin{array}{cc}
\rho_{11}+\rho_{22}&\rho_{13}+\rho_{24}\\
\rho_{31}+\rho_{42}&\rho_{33}+\rho_{44}
\end{array}\right)\right\}\nonumber\\
&&~~-\mbox{Tr}\left\{
\left(\begin{array}{cc}
\rho_{11}+\rho_{33}&\rho_{12}+\rho_{34}\\
\rho_{21}+\rho_{43}&\rho_{22}+\rho_{44}
\end{array}\right)\ln
\left(\begin{array}{cc}
\rho_{11}+\rho_{33}&\rho_{12}+\rho_{34}\\
\rho_{21}+\rho_{43}&\rho_{22}+\rho_{44}
\end{array}\right)\right\}.
\end{eqnarray}
If Tr\,$A=\mu=1$, the matrix $A$ can be interpreted either as the
density matrix of the two-qubit state or as the density matrix of
the qudit state with $j=3/2$. The density matrix must satisfy
inequality~(\ref{EX2}).

We can obtain an extra entropic inequality for the matrices created,
in view of the portrait map applied to the matrix $A$. For this, we
introduce the number $\widetilde N=N+2=6$ and construct the
6$\times$6 matrix $\widetilde A=\left(\begin{array}{ccc}
0&0&0\\
0&A&0\\0&0&0
\end{array}\right)$. We assume Tr$\,A=1$. Then, since $6=2\cdot 3$
and $6=3\cdot 2$, we obtain the two following inequalities:
\begin{eqnarray}
-\mbox{Tr}(A\ln A)&\leq& -\mbox{Tr}\left\{
\left(\begin{array}{cc}
\rho_{44}+\rho_{22}&\rho_{23}\\
\rho_{32}&\rho_{33}+\rho_{11}
\end{array}\right)\ln\left(\begin{array}{cc}
\rho_{44}+\rho_{22}&\rho_{23}\\
\rho_{32}&\rho_{33}+\rho_{11}
\end{array}\right)\right\}\nonumber\\
&&-\mbox{Tr}\left\{
\left(\begin{array}{ccc}
\rho_{11}&\rho_{13}&0\\
\rho_{31}&\rho_{22}+\rho_{33}&\rho_{24}\\
0&\rho_{42}&\rho_{44}\end{array}\right)\ln
\left(\begin{array}{ccc}
\rho_{11}&\rho_{13}&0\\
\rho_{31}&\rho_{22}+\rho_{33}&\rho_{24}\\
0&\rho_{42}&\rho_{44}\end{array}\right)\right\},\label{EX3}\\
-\mbox{Tr}(A\ln A)&\leq& -\mbox{Tr}\left\{
\left(\begin{array}{cc}
\rho_{11}+\rho_{22}&\rho_{14}\\
\rho_{41}&\rho_{33}+\rho_{44}
\end{array}\right)\ln\left(\begin{array}{cc}
\rho_{11}+\rho_{22}&\rho_{14}\\
\rho_{41}&\rho_{33}+\rho_{44}
\end{array}\right)\right\}\nonumber\\
&&-\mbox{Tr}\left\{
\left(\begin{array}{ccc}
\rho_{33}&\rho_{34}&0\\
\rho_{43}&\rho_{11}+\rho_{44}&\rho_{12}\\
0&\rho_{21}&\rho_{22}\end{array}\right)\ln
\left(\begin{array}{ccc}
\rho_{33}&\rho_{34}&0\\
\rho_{43}&\rho_{11}+\rho_{44}&\rho_{12}\\
0&\rho_{21}&\rho_{22}\end{array}\right)\right\}.\label{EX4}
\end{eqnarray}

Thus, we showed that the matrix $A$~(\ref{EX1}) with Tr$\,A=1$
satisfies entropic inequalities (\ref{EX2})--(\ref{EX4}).

In the case of two qubits, inequality~(\ref{EX2}) coincides with the
quantum subadditivity condition, i.e., with the entropic inequality
$S(1,2)\leq S(1)+S(2)$, where the left-hand side of (\ref{EX2}) is
equal to the von Neumann entropy of the two-qubit state, and both
terms in the right-hand side of (\ref{EX2}) are the entropies of the
first and second qubits, respectively. For qudit with $j=3/2$, the
inequality was discussed in
\cite{OlgaVovaVI,Ruskai1,Markovich2}.

For $N=3$, we can choose $\widetilde N=N+1=4$ and apply the obtained
inequality to an arbitrary Hermitian matrix
$A=\left(\begin{array}{ccc}
\rho_{11}&\rho_{12}&\rho_{13}\\
\rho_{21}&\rho_{22}&\rho_{23}\\
\rho_{31}&\rho_{32}&\rho_{33}
\end{array}\right)$ and the matrix $
\widetilde A=\left(\begin{array}{cc}
A&0\\
0&0\end{array}\right)$ with blocks $$a_{11}=
\left(\begin{array}{cc}
\rho_{11}&\rho_{12}\\
\rho_{21}&\rho_{22}\end{array}\right),\quad
a_{12}= \left(\begin{array}{cc}
\rho_{13}&0\\
\rho_{22}&0\end{array}\right),\quad
a_{21}=
\left(\begin{array}{cc}
\rho_{31}&\rho_{32}\\
0&0\end{array}\right),\quad a_{23}=
\left(\begin{array}{cc}
\rho_{33}&0\\
0&0\end{array}\right).$$ We arrive at the inequality
\begin{eqnarray}\label{EX5}
&&~~-\mbox{Tr}\left\{
\left(\begin{array}{cccc}
\rho_{11}+x&\rho_{12}&\rho_{13}&0\\
\rho_{21}&\rho_{22}+x&\rho_{23}&0\\
\rho_{31}&\rho_{32}&\rho_{33}+x&0\\
0&0&0&x
\end{array}\right)\ln
\left(\begin{array}{cccc}
\rho_{11}+x&\rho_{12}&\rho_{13}&0\\
\rho_{21}&\rho_{22}+x&\rho_{23}&0\\
\rho_{31}&\rho_{32}&\rho_{33}+x&0\\
0&0&0&x
\end{array}\right)\right\}\nonumber\\
&&~~-(\rho_{11}+\rho_{22}+\rho_{33}+4x)\ln(\rho_{11}+\rho_{22}+\rho_{33}+4x)\nonumber\\
&&\leq -\mbox{Tr}\left\{
\left(\begin{array}{cc}
\rho_{11}+\rho_{22}+2x&\rho_{13}\\
\rho_{31}&\rho_{33}+2x\\
\end{array}\right)\ln\left(\begin{array}{cc}
\rho_{11}+\rho_{22}+2x&\rho_{13}\\
\rho_{31}&\rho_{33}+2x\\
\end{array}\right)\right\}\nonumber\\
&&~~-\mbox{Tr}\left\{
\left(\begin{array}{cc}
\rho_{11}+\rho_{33}+2x&\rho_{12}\\
\rho_{21}&\rho_{22}+2x\\
\end{array}\right)\ln
\left(\begin{array}{cc}
\rho_{11}+\rho_{33}+2x&\rho_{12}\\
\rho_{21}&\rho_{22}+2x\\
\end{array}\right)\right\},
\end{eqnarray}
where $x$ is such a number that $A_0+x\geq 0$ and $A_0$ is the
minimum negative eigenvalue of the Hermitian matrix $A$. If the
3$\times$3 matrix $A$ is the density matrix of a qutrit state (we
assume $x=0$ and Tr$\,A=1$), inequality~(\ref{EX5}) coincides with
the subadditivity condition considered in \cite{OlgaVova1}.

\section{Conclusions}
\pst
To conclude, we point out the main results of our work.

We found the new matrix inequality valid for an arbitrary Hermitian
matrix; it is given by Eq.~(\ref{LM10}). We introduced the notion of
matrix information; see Eq.~(\ref{LM5}).

For arbitrary $N$$\times$$N$ matrix $A$, we constructed the matrix
portrait $\Phi(A)$, which is the linear map $A\to A'=\Phi(A)$, being
an analog of the partial tracing procedure used to obtain the matrix
factors $B$ and $C$ presenting the matrix $A$ in the form of tensor
product of these factors, $A=B\otimes C$. Employing the matrix
portraits $B$ and $C$ and embedding the matrices $A$, $B$, and $C$
in the linear space of higher dimensions, we obtained new entropic
matrix inequalities written for the Hermitian matrix $A$ in the
explicit form. Considering the subset of the set of matrices $A$,
which contains all density $N$$\times$$N$ matrices of the systems of
qudits, we derived new entropic matrix and information inequalities
for the density matrices.

Due to the procedure suggested here, we extended the known entropic
subadditivity condition for bipartite quantum systems to the case of
arbitrary single qudit systems. The method to obtain for a single
qudit state all entropic inequalities known for composite systems,
including the inequalities for the von Neumann entropy and
$q$-entropy~\cite{Renyi,Tsallis,Rastegin} along with the Bell-like
inequalities~\cite{SHRS,AkopyanJRLR,GisinArXiv}, can be formulated
as a straightforward continuation of the tools demonstrated in this
work.

We presented the map of the $N$$\times$$N$ matrix $A\to\Phi(A)$ in
the case of factorization $N=nm$. Repeating the map algorithm
step-by-step, one can construct a chain of maps for
$N=\prod_{k=1}^MN_k$, where $N_k$ are integers, and also in the case
where $\widetilde N=N+s=\prod_{k=1}^{\widetilde M}N_k$.

Since we understood that a single qudit state could have the
entanglement properties analogous to the entanglement properties of
multiqudit systems, we suggested to apply this knowledge to study
the resource of entanglement to be used for quantum computing,
analogously as it takes place in the case of composite quantum
systems~\cite{Nielson}.

The obtained map of the density matrix of a single qudit state on
the density matrix of a multiqudit state, including the $N$-qubit
state, provides the possibility to classify the quantum channels
transforming the separable states into entangled states, and vice
versa, of the single qudit. This possibility is related to the
identity of the $N$-dimensional Hilbert-space properties, which do
not depend on the interpretation of the Hilbert space as the space
of states of composite or noncomposite systems. Since there exists
the strong subadditivity condition for the density matrix of the
three-partite system~\cite{Ruakai1}, we can obtain a new matrix
inequality, which is an analog of this condition, for an arbitrary
Hermitian $N$$\times$$N$ matrix, including the density matrix of the
single qudit state. We continue the consideration of the found
matrix inequalities in the form of relations for qudit tomograms of
classical and quantum system
states~\cite{Beauty,RitaPS13,Elze,JRLR2013}, empolying the
inequalities for the probability vectors depending on the parameters
of the unitary matrix in a future publication.


\end{document}